\documentstyle[aps,prl,epsf]{revtex}
\begin{document}
\twocolumn[\hsize\textwidth\columnwidth\hsize\csname@twocolumnfalse\endcsname
\title {Josephson Plasma Resonance as a Structural Probe of Vortex
Liquid}
\author{A.E. Koshelev$^a$, L.N. Bulaevskii$^b$ and M.P. Maley$^b$}
\address{$^a$ Materials Science Division, Argonne National Laboratory, 
Argonne, Illinois 60439 \\
$^b$ Los Alamos National Laboratory, Los Alamos, NM 87545
}
\date{\today}
\maketitle
\begin{abstract}
Recent developments of the Josephson plasma resonance and transport 
c-axis measurements in layered high T$_{c}$ superconductors allow 
to probe Josephson coupling in a wide range of the vortex 
phase diagram.  We derive a relation between the field dependent 
Josephson coupling energy and the density correlation function 
of the vortex liquid.  This relation provides a unique opportunity to 
extract the density correlation function of pancake vortices from the 
dependence of the plasma resonance on the $ab$-component of the 
magnetic field at a fixed $c$-axis component.

\end{abstract}

]

\narrowtext

Thermal suppression of Josephson interlayer coupling by vortex 
fluctuations \cite{decoupl} is the key factor, which enhances vortex 
mobility and limits useful applications of Bi- and Tl-based high 
T$_{c}$ superconductors (HTSC).  Recent discovery of the theoretically 
predicted \cite{mish,bmt} Josephson plasma resonance (JPR) effect in 
the mixed state of highly anisotropic layered high temperature and 
organic superconductors \cite{oph,mats,shib,kadow,MatsAng,Bayrakci} 
provides a new and unique opportunity to probe the Josephson coupling 
of layers in a wide range of magnetic fields and temperatures.  An 
alternative way to probe Josephson coupling in the vortex state is 
direct $c$-axis transport measurements \cite{Transport}.

A minimum necessary condition for the existence of a narrow JPR line 
is short-time/short-range phase coherence between the superconducting 
layers.  In the London regime this coherence is quantitatively 
characterized by the spatial and temporal behavior of the ``local 
coherence parameter'' ${\cal C}_{n}({\bf r},t)=\cos[\varphi 
_{n,n+1}({\bf r},t)]$, with $\varphi _{n,n+1}=\varphi _{n+1}-\varphi 
_{n}-(2\pi s/\Phi _{0})A_{z}$ being the gauge-invariant phase 
difference between layers $n$ and $n+1$.  Here ${\bf r}=(x,y)$ and $z$ 
are coordinates in the $ab$ plane and along the $c$ axis, $s$ is the 
interlayer spacing.

The first important condition is that the typical time of variation of 
this factor (phase slip time) must be much larger than the period of 
plasma oscillations.  Their ratio was estimated \cite{kosh} as 
$\approx 10^4$ at $T=45$ K and magnetic field $B_{z}=0.1$ T. In this 
case the plasma mode probes a snapshot of the instantaneous phase 
distribution.

In the decoupled liquid state the cosine factor ${\cal C}_{n}({\bf 
r},t=0)$ also rapidly oscillates in space because correlations between 
pancake positions in neighboring layers are almost absent.  The 
possibility of a sharp resonance in the situation when fluctuations of 
local Josephson coupling are much stronger than its average is an 
interesting and nontrivial issue.  JPR resonance in such situation 
occurs because small phase oscillations induced by the external 
electric field change slowly in space and average out these rapid 
variations \cite{bdmb,LongPaper}.

In this Letter, using the high temperature expansion with respect to 
the Josephson coupling, we establish a relation connecting the plasma 
frequency with the density correlation function of the pancake liquid.  
A similar relation was derived in Ref.\ \cite{bpm} assuming the 
Gaussian distribution for the interlayer phase fluctuations.  We 
demonstrate that the pair distribution function of the liquid is 
unambiguously connected with the dependence of the plasma frequency on 
the in-plane component of the magnetic field for fixed $c$-axis 
component.  Therefore, this dependence can be used to extract 
quantitative information about the structure of the liquid phase.

The essential physics of JPR is captured by a simplified equation for 
small oscillations of the phase difference $\varphi_{n,n+1}^{\prime 
}({\bf r},\omega)$ induced by the external microwave electric field 
with the amplitude ${\cal D}_{z}$ and the frequency $\omega$ (see
Ref.~\onlinecite{bdmb})
\begin{equation}
\left[ \frac{\omega ^{2}}{\omega_{0}^{2}}+\lambda 
_{J}^{2}\hat{L}\nabla_{{\bf r}} ^{2}-{\cal C}_{n}({\bf r})\right] 
\varphi _{n,n+1}^{\prime }=-\frac{\hbar i\omega {\cal D}_{z}}{8\pi 
eE_{J}},
\label{DynEq}
\end{equation}
where $\omega _{0}(T)=c/\sqrt{\epsilon _{c}}\lambda _{c}(T)$ is the
zero field plasma frequency, $\epsilon_{c}$ is the high frequency
dielectric constant, $\lambda_{c}$ and $\lambda_{ab}$ are the
components of the London penetration depth, $E_{J}=E_{0}/\lambda_J^2$
is the Josephson energy per unit area, $E_{0}=s\Phi
_{0}^{2}/16\pi^{3}\lambda _{ab}^{2}$, $\lambda_{J}=\gamma s$ is the
Josephson length, and $\gamma=\lambda_{c}/\lambda_{ab}$ is the
anisotropy ratio.  The inductive matrix $\hat{L}$ is defined as
$\hat{L}A_{n}=\sum_{m}L_{nm}A_{m}$ with $ L_{nm}\approx (\lambda
_{ab}/2s)\exp\left( -|n-m|s/\lambda _{ab}\right)$.  We neglect in
Eq.~(\ref{DynEq}) time variations of ${\cal C}_{n}({\bf r},t)$
assuming them to be small during the time $1/\omega$.
Eq.~(\ref{DynEq}) also does not take into account charging effects
\cite{koy} and quasiparticle dissipation.  These effects don't 
influence much the position of the resonance but may modify its lineshape.  

Static configurations $\varphi_{n,n+1}({\bf r})$ are mainly determined 
by thermal fluctuations of pancake vortices and quantitatively 
characterized by the average value of the cosine factor, ${\cal 
C}({\bf B},T)$,
\begin{equation}
{\cal C}({\bf B},T)=\langle \cos \varphi _{n,n+1}({\bf r})\rangle  
\label{AvCos}
\end{equation}
and its static correlation function $S({\bf r},{\bf B},T)$, 
\begin{equation}
S({\bf r},{\bf B},T)=\langle \cos [\varphi _{n,n+1}({\bf 
r})-\varphi_{n,n+1}(0)]\rangle,
\label{coscorr}
\end{equation}
where $\langle \ldots \rangle $ denotes the thermal average.  In this 
Letter we consider the pancake liquid state where the 
correlations between pancake arrangements in different layers are 
almost absent.  In such situation ${\cal C}_{n}({\bf r})$ rapidly 
oscillates in space, so that $S({\bf r},{\bf B},T)$ drops at distances of
the order of the intervortex spacing $a=(\Phi_0/B_z)^{1/2}$, and ${\cal 
C}({\bf B},T)\ll 1$.

An important observation is that in spite of rapid variations of 
${\cal C}_{n}({\bf r})$ the oscillating phase $\varphi^{\prime } 
\equiv \varphi_{n,n+1}^{\prime }({\bf r})$ varies smoothly in space.  The
typical length scale $L_{\varphi}$ of $\varphi^{\prime }$
variations can be 
estimated by balancing the typical kinetic energy of supercurrents, 
$E_{0}(\varphi^{\prime })^{2}/L_{\varphi}^{2}$, with the typical 
Josephson energy, $E_{J}(\varphi^{\prime })^{2} a/L_{\varphi}$, in the 
vortex state which is strongly disordered along the $c$ axis.  This 
gives $L_{\varphi}=\lambda_{J}^{2}/a$.  Smoothly varying 
$\varphi^{\prime }$ effectively averages rapid variation of ${\cal 
C}_{n}({\bf r})$ and the plasma frequency is simply determined by 
${\cal C}({\bf B},T)$ (see Refs.~\onlinecite{bmt,kosh})
\begin{equation}
\omega _{p}^{2}({\bf B},T) =\omega_{0}^{2}(T){\cal C}({\bf B},T)/{\cal
C}(T),  
\label{omegap} 
\end{equation}
Here the factor ${\cal C}(T)={\cal C}(0,T)$ takes into account the 
suppression of zero field plasma frequency by phase fluctuations 
(see below).  Fluctuations of ${\cal C}_{n}({\bf r})$ smoothened over 
the area $L_{\varphi}^2$, ${\cal 
C}(L_{\varphi})=L_{\varphi}^{-2}\int_{r<L_{\varphi}}d{\bf r}{\cal 
C}_n({\bf r})$, produce inhomogeneous broadening of the JPR line.  
Calculating the mean squared fluctuation of ${\cal C}(L_{\varphi})$, 
$\langle[{\cal C}(L_{\varphi})-{\cal C}]^{2}\rangle\approx 
a^{2}/L_{\varphi}^{2}=a^{4}/\lambda_{J}^{4}$, we obtain the estimate for 
the inhomogeneous line width
\begin{equation}
\delta(\omega_p^2)\approx\omega_0^2(T)a^2/\lambda_J^2. 
\label{line}
\end{equation}
Detailed calculations of the intrinsic lineshape due to this mechanism 
will be published elsewhere \cite{LongPaper}.

We establish now a relation between plasma frequency and the 
vortex liquid structure.  In general, the Josephson coupled system
in the London regime, $B_{z}\ll H_{c2}$, may be described in terms of 
vortex coordinates ${\bf r}_{n\nu}$ (index $\nu$ labels vortices in the 
layer $n$) and by the regular (spin-wave type) phase difference 
$\varphi _{n,n+1}^{(r)}({\bf r})$.  The free energy functional in 
terms of these variables is \cite{bdmb}
\begin{eqnarray}
{\cal F}\{{\bf r}_{n\nu },\varphi _{n,n+1}^{(r)}\}&=&\nonumber \\
{\cal F}_{v}({\bf r}_{n\nu })&+&{\cal F}_\varphi \{\varphi
_{n,n+1}^{(r)}\}+ 
{\cal F}_J \{{\bf r}_{n\nu },\varphi _{n,n+1}^{(r)}\}.  
\label{mainf}
\end{eqnarray}
Here ${\cal F}_{v}({\bf r}_{n\nu })$ accounts for the two-dimensional 
energy of pancakes and also their electromagnetic interaction in 
different layers \cite{clem}.  The second term,
\begin{equation}
{\cal F}_\varphi\{\varphi _{n,n+1}^{(r)}\}= 
\frac{E_0}{2}\sum_n\int d{\bf r}\nabla \varphi _{n,n+1}^{(r)}\hat{L}
\nabla \varphi _{n,n+1}^{(r)} ,
\label{vfunc}
\end{equation}
is the energy of intralayer currents associated with regular phase  
fluctuations, and the third term, 
\begin{eqnarray}
&&{\cal F}_J\{{\bf r}_{n\nu },\varphi _{n,n+1}^{(r)}\}= \nonumber \\
&&E_J\sum_n\int d{\bf r}\left[1-\cos 
\left(\varphi _{n,n+1}^{(v)}+\varphi _{n,n+1}^{(r)}-{\frac{2\pi 
s}{\Phi_{0}}}B_{x}y\right)\right],  
\label{FJ}
\end{eqnarray}
is the Josephson energy.  We split the phase difference $\varphi_{n,n+1}$
into vortex part, regular spin-wave part, and contribution coming from 
the in-plane magnetic field $B_{x}$,
\begin{equation}
\varphi_{n,n+1}({\bf r})=\varphi _{n,n+1}^{(v)}({\bf r})+\varphi
_{n,n+1}^{(r)}({\bf r})
-(2\pi s/\Phi_{0})B_{x}y. 
\end{equation}
The phase 
$\varphi_{n,n+1}^{(v)}({\bf r};{\bf r}_{n\nu},{\bf r}_{n+1,\nu})$ is the
singular part 
of the phase difference induced by vortices at positions ${\bf 
r}_{n\nu}$, ${\bf r}_{n+1,\nu}$ when Josephson coupling is absent
($E_J=0$):
\begin{eqnarray}
&&\varphi_{n,n+1}^{(v)}({\bf r};{\bf r}_{n\nu},{\bf r}_{n+1,\nu})=
\nonumber \\
&&\sum_{\nu}[\phi_v({\bf 
r}-{\bf r}_{n\nu})-\phi_v({\bf r}-{\bf r}_{n+1,\nu})],
\label{2D}
\end{eqnarray}
where $\phi_v({\bf r})$ is the polar angle of the point ${\bf r}$. 

We calculate ${\cal C}({\bf B},T)$ with the functional (\ref{mainf}) 
using high temperature expansion \cite{kosh} with respect to ${\cal 
F}_J$ as:
\begin{equation}
{\cal C}({\bf B},T)\approx \frac{E_{J}}{2T}\int d{\bf r}S({\bf
r},{\bf B},T)=f({\bf B},T)\frac{E_0B_J}{2TB_z}.
\label{cfun}
\end{equation}
Here $f({\bf B},T)=a^{-2}\int d{\bf r}S({\bf r},{\bf B},T)$ is a universal 
function and $B_{J}=\Phi_{0}/\lambda_{J}^{2}$.  Eq.~(\ref{cfun}) is 
valid until ${\cal C}({\bf B},T)\ll 1$ which corresponds to the field 
and temperature range $TB_{z}\gg E_0 B_J$. Comparing 
Eqs.~(\ref{omegap}) and (\ref{cfun}) with Eq.~(\ref{line}) we obtain 
an estimate for the relative line width in the liquid state due to the 
inhomogeneous broadening, $\delta\omega_p/\omega_p\approx 
T/E_0\ll 1$.

In the lowest order in $E_{J}$ spin-wave and vortex degrees of freedom 
do not interact:
\begin{equation}
S({\bf r},{\bf B})=S_{v}({\bf r},B_z)S_{r}({\bf r})\cos (2\pi
sB_{x}y/\Phi _{0}),  
\label{SvSr}
\end{equation}
where the correlation function $S_{v}({\bf r})$ is determined by the 
functional ${\cal F}_v$, while $S_r({\bf r})$ is determined by ${\cal 
F}_\varphi$.  In the pancake liquid regime, which we consider in this 
Letter, $B_{x}$ penetrates almost freely into the sample and has no 
influence on the phase fluctuations $\varphi _{n,n+1}^{(v)}$ and 
$\varphi_{n,n+1}^{(r)}$.  An extra spatial dependence of 
$\varphi_{n,n+1}$ induced by $B_{x}$ gives possibility to probe phase 
correlations at given $B_{z}$ by measuring $B_{x}$-dependence of the 
plasma frequency.

We now express the vortex phase correlation function $S_{v}({\bf r})$ 
via the density correlation function of pancake vortices 
induced by $B_{z}$.  The difference $\Phi_{v}({\bf 
r})=\varphi_{n,n+1}^{(v)}({\bf r})-\varphi_{n,n+1}^{(v)}(0)$ can be 
connected with pancake densities $\rho _{n}({\bf R})=\sum_{\nu }\delta 
({\bf R}-{\bf r}_{n\nu })$ in the layers as
\begin{equation}
\Phi _{v}({\bf r})=\int d{\bf R}[\rho _{n}({\bf R})-\rho _{n+1}({\bf R}
)]\beta ({\bf r},{\bf R}).  \label{Ph}
\end{equation}
Here $\beta ({\bf r},{\bf R})=\phi_v({\bf r}/2-{\bf R})-\phi_v(-{\bf 
r}/2-{\bf R})$ is the angle at which the segment connecting points 
$-{\bf r}/2$ and ${\bf r}/2$ is seen from the point ${\bf R}$, 
\[
\cos \beta (r,{\bf R})=(R^{2}-r^{2}/4)[(R^{2}+r^{2}/4)^{2}-({\bf 
Rr})^{2}]^{-1/2}.
\] 
The function $\beta ( {\bf r},{\bf R})$ has a jump of $2\pi $ when 
point ${\bf R}$ intersects segment $ [-{\bf r}/2,{\bf r}/2]$.  Using 
the Gaussian approximation for $\Phi_{v}({\bf r})$ we obtain for 
vortex phase correlation function $S_{v}({\bf r})\equiv \langle \cos 
[\Phi _{v}({\bf r})]\rangle$
\begin{equation}
S_{v}({\bf r})=\exp [-F_{v}(r)] 
\label{SvFv}
\end{equation}
with 
\begin{equation}
F_{v}(r)=\langle [\Phi _{v}({\bf r})]^{2}\rangle/2. 
\label{appr}
\end{equation}
Using Eqs.\ (\ref{Ph}) and (\ref{appr}) we connect $F_{v}(r)$ with the 
pair distribution function $h(r)$ defined by the relation
\begin{equation}
n_{v}^{2}h(r)=\langle \rho _{n}(0)\rho _{n}({\bf r})\rangle -\langle \rho
_{n}(0)\rho _{n+1}({\bf r})\rangle -n_{v}\delta ({\bf r})
\end{equation}
as 
\begin{eqnarray}
&&F_{v}(r)= \label{fvr} \\
&&-\frac{n_{v}^{2}}{2}\int d{\bf R}d{\bf r}^{\prime }h({\bf
r}^{\prime})\left[ 
\beta\left ({\bf r},{\bf R}+\frac{{\bf r}^{\prime 
}}{2}\right)-\beta\left ({\bf r},{\bf R}- \frac{{\bf r}^{\prime 
}}{2}\right)\right] ^{2}.  \nonumber
\end{eqnarray}
For this identity $n_{v}\int d{\bf r}h(r)=-1$ has been used, 
$n_{v}=\langle \rho _{n}( {\bf r})\rangle=a^{-2}$.  Performing 
integration with respect to ${\bf R}$ and averaging over directions of 
${\bf r}^{\prime}$ we finally obtain
\begin{equation}
F_{v}(r)=-\pi n_{v}^{2}\int r^{\prime }dr^{\prime }h(r^{\prime
})J(r,r^{\prime }),  \label{meq}
\end{equation}
where the universal function $J(r,r^{\prime })$ is given by 
\begin{eqnarray}
&&J(r,r^{\prime })=2\pi r^{2}\left[ \ln (r^{\prime }/r)+2\right] ,\ \ \
r^{\prime }>r,  \label{Jr} \\
&&J(r,r^{\prime })=4\pi \left[ 2rr^{\prime }-r^{\prime 2}-(r^{\prime
2}/2)\ln (r/r^{\prime })\right] ,\ \ \ r^{\prime }<r.  \nonumber
\end{eqnarray}
Eqs.~(\ref{SvFv}), (\ref{meq}) and (\ref{Jr}) represent general 
relations connecting phase fluctuations with vortex density 
fluctuations.  The function $F_{v}(r)$ has the following asymptotics
\begin{eqnarray*}
F_{v}(r) &\approx &-8\pi ^{2}\frac{r}{a}\int_{0}^{\infty 
}dxx^{2}h(x),\ \ \ r\gg a, \\
F_{v}(r) &\approx &  \frac{\pi r^{2}}{a^{2}} \left( \ln \frac{a}{r}
+2-2\pi\int_{0}^{\infty }dxxh(x)\ln x\right) ,\ \ r\ll a.
\end{eqnarray*}
Linear increase of $F_{v}(r)$ at large $r$ indicates 
exponential decay of phase correlations at large distances.

The problem of calculating of $S_v({\bf r})$ is now reduced to 
calculation of the integral (\ref{meq}) with the density correlation 
function of the liquid.  This function is not available 
analytically.  It can be calculated from Monte Carlo simulations or 
(approximately) using the density functional theory \cite{Menon}.  In 
the decoupled pancake liquid regime the correlations between two 
dimensional pancake liquids in different layers are very weak because 
the energy of the magnetic interlayer interaction of disordered 
pancakes has an additional small parameter $s^2/\lambda_{ab}^2$ in 
comparison with the intralayer energy.  If these correlations are 
neglected then the vortex system is equivalent to the one-component 
two-dimensional Coulomb plasma which was studied extensively 
\cite{caillol}.  The pair distribution function 
$h_{2D}(r,B_{z},T,E_{0})$ within this approximation has an exact 
scaling property $ h_{2D}(r,B_{z},T,E_{0})=h_{2D}(r/a,T/E_{0})$, i.e., 
it depends on magnetic field only through the spatial scale.  As 
follows from Eq.\ (\ref{meq}) this scaling property is also 
transferred to the function $F_{v}(r)$, $ 
F_{v}(r)=F_{v}(r/a,T/E_{0})$.  If, in addition, fluctuations of the 
regular phase are neglected then at $B_{x}=0$ the function $ 
f(T/E_{0})$ in the expression (\ref{cfun}) for ${\cal C}({\bf B},T)$ 
is field independent.  Therefore, the scaling property ${\cal 
C}(B_{z},T)\propto 1/B_{z}$ is a consequence of the approximation of 
completely decoupled pancake liquids in pinning-free layers when 
spin-wave phase fluctuations are negligible.

We generated the pair distribution functions $h(x)$ with $x=r/a$ for 
the two-dimensional pancake liquid using Langevin dynamics 
simulations.  The model and the algorithm have been described in 
Ref.~\onlinecite{kv}.  Fig.~1 shows an example of the pair 
distribution function $h(x)$ for temperature $T=0.012\pi E_0$ 
($\approx 1.7\ T_{m}$) and the corresponding function $F_v(x)$ 
obtained by numerical integration of Eq.~(\ref{meq}).  To show a weak 
waving of $F_v(x)$ due to liquid correlations we also plot the 
derivative of $F_v(x)$.
 
As temperature approaches $T_{c}$ the role of fluctuations of the regular
phase progressively increases.  The spin-wave phase correlations decay 
algebraically
\begin{equation}
S_{r}({\bf r})=(\xi_{ab}/r)^{2\alpha },\ \ \alpha =T/2\pi E_{0}(T),
\label{scff}
\end{equation}
where $\xi_{ab}(T)$ is the superconducting correlation length 
[$1/\xi_{ab}$ determines the upper cut-off for momenta characterizing 
spatial variations of $\varphi_{n,n+1}^{(r)}({\bf r})$ in 
Eq.~(\ref{vfunc})].  As follows from Eq.\ (\ref{SvSr}) the 
contribution from regular phase fluctuations becomes essential if 
$S_{r}(r\sim a)\ll 1 $.  This gives the condition $T>E_{0}(T)\ln(H_{c2}/B)$.
 
To relate $\omega_{p}(B,T)$ with the plasma 
frequency at zero field and the Josephson coupling $E_{J}\propto 
\lambda_{c}^{-2}$ we have to take into account that these quantities 
are also renormalized by the phase fluctuations.  Their 
suppression is determined by the cosine factor at $B=0$, $C(T)$, 
which was estimated in Ref.~\onlinecite{CritLay} as 
\begin{equation}
C(T)=(\xi_{ab}/\lambda_{J})^{\alpha}.
\label{CT}
\end{equation}

Finally using Eqs.~(\ref{omegap}), (\ref{SvSr}), 
(\ref{SvFv}), (\ref{scff}), and (\ref{CT}) we obtain an expression for
plasma frequency which 
incorporates effects of vortices and of regular phase fluctuations at 
all temperatures :
\begin{eqnarray}
&&\frac{\omega_{p}^{2}(B,T)}{\omega_{0}^{2}(T)}=\frac{E_{0}(T)}{2T }
\left(\frac{B_{J}}{B_{z}}\right)^{1-\alpha }
f_{s}\left( \frac{2\pi sB_{x}}{ \sqrt{\Phi 
_{0}B_{z}}}\right) , \label{result} \\
&&f_{s}(b)=2\pi \int_{0}^{\infty }dxx^{1-2\alpha }\exp
[-F_{v}(x)]J_{0}(bx),
\label{int}
\end{eqnarray}
where $J_{0}(x)$ is the Bessel function.  Note that regular phase 
fluctuations reduce the power index in the $B_{z}$ dependence of 
${\cal C}(B_{z},T)$ at $B_{x}=0$ and make it temperature dependent. 
Using the typical parameters for optimally doped 
Bi$_{2}$Sr$_{2}$CaCu$_{2}$O$_{8-x}$ [$\lambda_{ab}(0)=2000$ \AA, 
$T_{c}=90$ K] we obtain that the index $\alpha$ increases from 
$\alpha\approx 0.05$ at $T=50$ K to $\alpha\approx 0.18$ at $T=$75 K.

Measurements of $B_{x} $-dependence of the resonance frequency $\omega 
_{p}$ at fixed $B_{z}$ and $T$ allow one to obtain, in principle, the 
pair distribution function $h(x,T)$, i.e., to extract quantitative 
information about correlations in the vortex liquid.  Using 
Eq.~(\ref{result}) the function $f(b)$ can be found from the 
dependence $\omega _{p}(B_{x},B_{z})$ as:
\begin{equation}
f_{s}\left( \frac{2\pi sB_{x}}{\sqrt{\Phi _{0}B_{z}}}\right)
=\frac{\omega
_{p}^{2}(B_{x},B_{z})}{\omega _{p}^{2}(0,B_{z})}f_{s}(0).
\end{equation}
With $f_s(b)$ known, the function $F_{v}(x)$ may be found by the 
reverse Fourier transform from Eq.~(\ref{int}),
\begin{equation}
F_{v}(x)=-\ln \left[ x^{2\alpha }\int_{0}^{\infty
}bdbf_{s}(b)J_{0}(bx)\right] .
\end{equation}
Finally, the pair distribution function $h(x)$ can be expressed 
through $F_{v}(x)$ using Eqs.~(\ref{meq}) and (\ref{Jr}):
\begin{equation}
h(x)=-\frac{1}{8\pi x^{3}}\frac{d}{dx}\left\{ x^{3}\frac{d}{dx}\left[ 
x\frac{ d^{3}F_{v}(x)}{dx^{3}}\right] \right\} .  \label{de}
\end{equation}
Unfortunately, high order differentiations in the last equation 
complicate practical realization of the procedure.  As it is shown in 
Fig.~1, oscillating behavior of liquid correlations is hardly seen in 
$F_v(x)$ but the first derivative, $F_v'(x)$, may expose it quite 
clearly.  The experimental angular dependence of the plasma frequency 
\cite{MatsAng,Bayrakci} was found to be in a good qualitative 
agreement with Eq.~(\ref{result}).  However an accuracy of existing 
measurements is not sufficient to perform the described quantitative 
analyses.

In conclusion, we obtain a general expression connecting the plasma 
frequency with the density correlation function of pancake 
liquid.  We demonstrate that the dependence of the plasma frequency on 
the magnetic field component parallel to the layers $B_x$ at fixed 
perpendicular component $B_z$ contains full structural information 
about the vortex liquid at given $B_z$ and temperature.

This work was supported by the National Science Foundation Office of 
the Science and Technology Center under contract No.  DMR-91-20000  
and by the U. S. Department of Energy, BES-Materials Sciences, under 
contract No.  W-31- 109-ENG-38.  Work in Los Alamos is supported by 
the U.S. DOE.

\begin{figure}
\epsfxsize=3.4in \epsffile{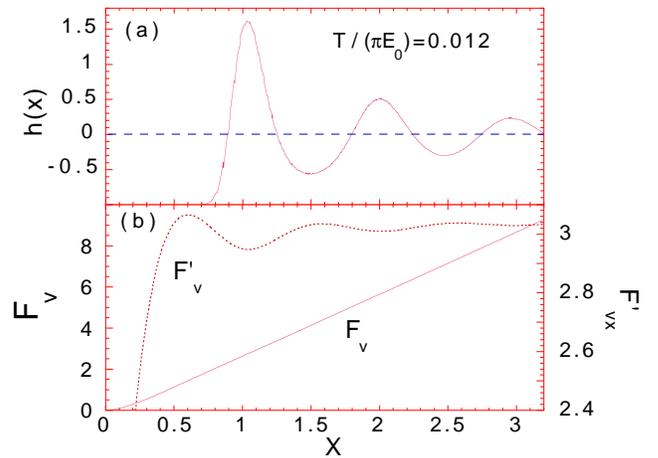} 
\caption{(a) Pair distribution 
function $h(x)$ obtained by Langevin dynamics simulations of two 
dimensional vortex liquid for $T/\pi E_{0}=0.012$; (b) Plot of mean 
squared phase difference between points separated by distance $r=x a$, 
$F_{v}(x)$, [see Eq.\ (\protect\ref{appr})] obtained by numerical 
integration of Eq.\ (\protect\ref{meq}) with plotted above $h(x)$.  
Derivative of $F_{v}(x)$ is also plotted to enhance weak oscillations 
due to liquid correlations.}
\label{Fig-hFv}
\end{figure}
\end{document}